\documentstyle[12pt,psfig]{article}

\def\xray{$X$--ray  }

\def\n_med{{\left<n\right>}}

\def\begc{\begin{center} }
\def\endc{\end{center} }
\def\begf{\begin{figure} }
\def\endf{\end{figure} }

\def\j3{{J_3}}

\newcommand{\mincir}{\raise -2.truept\hbox{\rlap{\hbox{$\sim$}}\raise5.truept
\hbox{$<$}\ }}
\newcommand{\magcir}{\raise -2.truept\hbox{\rlap{\hbox{$\sim$}}\raise5.truept
\hbox{$>$}\ }}
\newcommand{\siml}{\raise -2.truept\hbox{\rlap{\hbox{$\sim$}}\raise5.truept
\hbox{$<$}\ }}
\newcommand{\simg}{\raise -2.truept\hbox{\rlap{\hbox{$\sim$}}\raise5.truept
\hbox{$>$}\ }}
\newcommand{\be}{\begin{equation}}
\newcommand{\ee}{\end{equation}}
\newcommand{\ba}{\begin{eqnarray}}
\newcommand{\ea}{\end{eqnarray}}
\newcommand{\brr}{\begin{array}}
\newcommand{\err}{\end{array}}
\newcommand{\bc}{\begin{center}}
\newcommand{\ec}{\end{center}}

\newcommand{\hm}{\,h^{-1}{\rm Mpc}}
\newcommand{\vel}{\,{\rm km\,s^{-1}}}
\newcommand{\fl}{\,{\rm erg\,s^{-1}cm^{-2}}}
\newcommand{\lum}{\,{\rm erg\, s^{-1}}}

\begin{document}

\sf 
\baselineskip=18pt

\title{X-ray Clusters of Galaxies \\as Tracers of Structure in the
Universe} 
\author{Stefano Borgani$^1$ \& Luigi Guzzo$^2$ \\
\\
$^1$ INFN, Sezione di Trieste \\
c/o Dipartimento di Astronomia dell'Universit\`a, \\
Via Tiepolo 11, I-34131, Trieste, Italy\footnote{also, 
INFN, Sezione di Perugia, Perugia, Italy}\\
borgani@ts.astro.it\\
$^2$ Osservatorio Astronomico di Brera \\
Via Bianchi 46, I-23807, Merate, Italy\\
guzzo@merate.mi.astro.it}
\date{\sl{To appear as a review in Nature\\ 4 January 2001 issue}}
\maketitle


{\bf Clusters of galaxies outline the network of the distribution of
visible matter in the Universe, marking the highest--mass knots where
filamentary structures join together.  If we observe the sky in $X$
rays, clusters of galaxies stand out as cosmic lighthouses by virtue
of a thin gas trapped and heated within their gravitational potential
wells.  This powerful emission is directly linked to the total
gravitating mass they contain, such that they can be efficiently used
as tracers of the cosmic mass distribution within a sizeable fraction
of the observable Universe. Recent observational campaigns have for
the first time used $X$--ray clusters to map cosmic structures on
scales approaching 3 billion light years. The emerging picture is
remarkably consistent with the expectation of a low--density Universe
dominated by cold dark matter.}

The present--day appearance of the galaxy distribution is the result
of the gravitational growth of the initial fluctuations laid down
during the very early stages of cosmic history, coupled to dissipative
processes which lit up matter and made it visible to our telescopes.
Cosmological models directly predict the distribution and
gravitational growth of the mass, not of the light we actually observe
from galaxies.  Thus, the grand challenge of unveiling the nature of
cosmic initial conditions from the observed structure of the Universe
is a process which is ideally characterised by two steps. The first
involves constructing maps of the distribution of luminous objects on
sufficiently large scales, so as to encompass a 
representative portion of the whole Universe.  The second requires
relating such ``light maps'' to the underlying mass through a
physically motivated and
robust recipe, the two distributions being in principle not
related by a one--to--one correspondence.  Although nowadays galaxy
surveys are able to probe the distribution of luminous structures over
scales larger than 100 Megaparsecs ({\sf Mpc} , 1 pc=3.26 light
years), their relation to the actual mass distribution depends on the
still poorly understood physics of galaxy formation and evolution.

Clusters of galaxies, the most evident concentrations in galaxy maps,
can themselves be used as tracers of the large--scale structure of the
Universe.  While missing the fine details, they can be efficiently
used to study extremely large volumes at a reduced cost in terms of
telescope time with respect to fully--sampled galaxy surveys.  Already
the earliest statistical samples of clusters from visually--compiled
catalogues \cite{bi:A58,bi:ACO89,Zwicky} reached typical depths of few
hundreds Mpc, and the Abell catalogue in particular \cite{bi:A58}
represents still today one of the main resources for cosmological
studies (e.g. \cite{Retzlaff,Miller}).  It is within the
``gravitational sinks'' of galaxy clusters that evidence for the
elusive dark matter was first found in the thirties
\cite{ZwickyDM,Smith36}, as a necessary ingredient to explain the fast
motions observed for cluster galaxies.

The loose definition of a cluster as a collection of galaxies is
however intrinsically uncertain, definitely not optimal for estimating
its mass, as required for linking observations and theoretical
predictions. Fortunately, about 20--30\% of the optically invisible
mass of a cluster is in the form of a diffuse hot gas \cite{bi:evr97},
trapped and heated to a temperature of the order of ${10^8}$ K by its
gravitational potential. At such high temperatures, this gas is a
fully ionised plasma, producing a powerful $X$--ray emission by
free--free electron--ion interactions, the so--called {\sl bremsstrahlung}
radiation. With total luminosities of $\sim 10^{43}-10^{45}\,{\rm
erg\, s^{-1}}$, and large physical dimensions ($\sim 1\,{\rm Mpc}$),
galaxy clusters can be recognised over the rather sparse $X$--ray sky
\cite{Hasinger} as extended sources, out to very large distances.  The
$X$--ray luminosity is shown to correlate well with the cluster mass
and indeed provides a fairly direct way to make a robust comparison of
the observed clustering with the predictions of cosmological models.
The most recent projects finalised to study large--scale structure
benefiting of these advantages
\cite{bi:xbacs,bi:bcs,bi:rass,Hans98,bi:noras}, are based on the
all-sky survey by the {\tt ROSAT} satellite and have so far
accumulated distances for $\sim 1000$ clusters within a volume with
size of the order of 1000 Mpc. The degree of clumpiness observed in
these samples on scales of several hundreds Mpc is remarkable, and
larger than previously indicated by galaxy surveys.  Very
interestingly, the amplitude and shape of the corresponding
distribution of mass inhomogeneities, which for $X$--ray clusters can
be precisely computed, point toward the picture of a Universe
dominated by non--baryonic cold dark matter, whose density is about
one third of that necessary for it to recollapse under its own
self--gravity.  Coupled to the recent strong evidence for a nearly
flat spatial geometry from the anisotropy of the Cosmic Microwave
Background \cite{bi:boom,bi:max}, this reinforces the apparent need
for an extra cosmic energy, what is commonly associated with the
cosmological constant.

\section*{\sf Large--scale view of the Universe}
Modern cosmology uses the distribution of galaxies and clusters to map
the structure of the Universe on scales which have presently reached a
few $100\,h^{-1}{\rm Mpc}$ (here $h$ is the Hubble constant, $H_0$, in
units of 100 km s$^{-1}$ Mpc$^{-1}$; see Box 1).  First systematic
{\sl redshift surveys} of galaxies started around the half of the
seventies \cite{rood} and became an industry during the last two
decades (see ref. \cite{bi:texas} and references therein; also, see
Box 1 for the definition of redshift). The main features of the
large--scale distribution of galaxies emerging from these cosmic maps
include long, thin superclusters, with stronger concentrations -- rich
clusters -- located at their intersections, surrounding large regions
essentially devoid of galaxies.  This is explicitly illustrated by the
top panel of Fig.~1, where we have plotted the distribution of about
26,000 galaxies composing the Las Campanas Redshift Survey (LCRS,
blue points) \cite{lcrs}.  The evident inhomogeneity in the
distribution of galaxies can be statistically quantified at the
simplest level in terms of their {\sl two--point correlation function}
$\xi(r)$, which measures the excess probability with respect to a
random distribution, to observe a pair of galaxies separated by a
distance $r$ \cite{peebles1993} (see Box 1).  The squares in Fig.~2
shows the correlation function measured from the LCRS galaxy survey
\cite{xi_lcrs}.  Consistently with other samples of normal galaxies,
$\xi(r)$ is well described by a power--law $(r/r_o)^{-\gamma}$, with
$r_o\simeq 5\,h^{-1}{\rm Mpc}$ and $\gamma\simeq 1.8$ for separations
$r$ between about 0.1 and $10\hm$.  This roughly tells us that
galaxies are strongly clustered within these two decades of scales,
while tending to a more homogeneous distribution at larger separations
\cite{Stef_PR,Gigi_NA,Wu_hom}.

This variety of observed cosmic structures is interpreted as arising
from the gravitational growth of initially tiny fluctuations,
generated in the early stages of the life of the Universe
\cite{bi:JAP_book,bi:CL95_book}. Cosmological models yield predictions
for the character of such initial fluctuations in terms of their {\sl
power spectrum} $P(k)$, i.e. the Fourier transform of $\xi(r)$ (see
Fig.~3 and Box 1).  However, their connection to the observations is
not trivial.  While direct and reliable predictions can be made about
the distribution of matter concentrations of a given mass
\cite{bi:MW96}, it is far less straightforward to describe the
clustering properties of the galaxies which have then formed within
these potential wells.  A suitable recipe to relate their observable
characteristics (as luminosity, colour and morphology) to their actual
total mass, would in fact require a complete understanding of all
physical processes regulating the formation and evolution of stars
within galaxies.  Although intensive work is currently dedicated to a
better comprehension of such mechanisms (e.g.,
\cite{bi:SP99,bi:Bens00} and references therein), this is far from
being satisfactory.

\section*{\sf Clusters as tracers of the cosmic web}
Once we accept the idea of luminous objects as tracers of the
underlying mass structure of the Universe, we might think to
alternatives to using single galaxies for this scope.  The bottom
panel in Fig.~1 plots the distribution of a large sample of clusters
of galaxies\cite{Hans98,Gigi99} detected by the X-ray satellite {\tt
ROSAT}, one typical example of which is shown in the combined
optical--$X$--ray picture of Fig.~4.  Comparison of the two panels of
Fig.~1 gives an idea of how massive clusters provide a coarser mapping
of large--scale structure, but are very efficient to sample extremely
large ($r > 100\hm$) volumes, which translates into a more modest
investment of telescope time to reconstruct their 3D distribution.  In
general, rich clusters as that of Fig.~4 contain several hundreds of
galaxies within a typical size of $\sim 1\,h^{-1}$ Mpc, and represent
the largest gravitationally bound structures in the Universe.
Galaxies in clusters are observed to move along their orbits with a
typical line--of--sight velocity dispersion $\sigma_v \sim $
1000$\vel$, so that the time required to a galaxy to cross the cluster
is approximately $t_{cr}={r/\sigma_v}\sim 10^9$ yr.  This means that,
given the typical age of the Universe, $t_U\simeq H_0^{-1}\simeq
10^{10}\,h^{-1}$ yrs, rich clusters had enough time to evolve into
dynamically relaxed systems.  While a continuous infall of galaxies
moving along radial orbits takes place in the cluster outskirts, the
central regions have therefore reached virial equilibrium, in which
the total mass is related to the galaxy velocity dispersion as
$M_{vir}\simeq 3\sigma_v^2 r / G\sim 10^{15}\,h^{-1}M_\odot$.  Already
first pioneering attempts \cite{ZwickyDM,Smith36} showed that cluster
masses measured in this way are about ten--times larger than expected
from the sum of the masses of the individual member galaxies.  This
observation represented the earliest evidence for a ``hidden'' form of
matter permeating the Universe.

In Figs.~2 and 3 we have also plotted the correlation function and
power spectrum measured for clusters selected through their $X$--ray
emission \cite{xi_REFLEX,pk_REFLEX}.  Both figures show how clusters
display a clustering amplitude significantly larger than galaxies,
while maintaining a similar shape: the correlation length of $X$--ray
luminous clusters is roughly 4 times larger, $r_o\simeq 20\,h^{-1}{\rm
Mpc}$. This gives us an explicit demonstration of how changing the
kind of tracer we are using, we measure different clustering
properties.  When first observed in optically--selected samples
\cite{BS83,KK83,Post98} this motivated the concept of {\sl bias} in
the distribution of cosmic structures \cite{bi:K84}: the strong
clumpiness of the cluster distribution is just the natural consequence
of clusters tracing only the high--density peaks of the underlying
mass density field.  More in general, in the cosmologists' jargon, the
word ``bias'' is meant to indicate the relation between the
distribution of a given class of observable objects, like galaxies and
clusters, and the underlying distribution of matter: {\sl what
regulates the amount of clustering of a class of objects is just the
characteristic mass of the objects themselves}.  In the case of
clusters, their correlation function and power spectrum are predicted
to be amplified with respect to those of the matter distribution, by a
constant {\sl biasing} factor, whose value is uniquely known once the
cluster mass and the cosmological model are specified \cite{bi:MW96}.
This has the important consequence that, given a sample of clusters
selected according to their mass, the observed clustering is a
truly direct test of the cosmological model.  Surprising as it might
seem to the reader, despite clusters are evidently more ``biased''
tracers of the mass distribution with respect to galaxies, they are
more effective because their biasing factor can be more easily computed.

Such a theoretical simplicity has however to face the somewhat loose
visual definition of a cluster as an agglomerate of galaxies.  The
traditional way of selecting clusters is in fact based on the
``eyeball'' detection of overdensities in the projected galaxy
distribution on the sky \cite{bi:A58}. In the attempt to provide a
simple estimator of the cluster mass, a {\sl richness} is measured by
counting, within a fiducial aperture radius, the number of galaxies
which belong to the cluster.  However, partly for the difficulty of
properly correcting for the contamination by background and foreground
galaxy counts \cite{bi:Lum92,bi:Eke_cl}, the richness itself is
intrinsically a poor mass indicator when compared to $X$--ray
luminosity (see Fig.~5). This highlights the difficulty of using the
distribution of clusters selected by optical richness to
quantitatively constrain theoretical scenarios for the formation of
cosmic structures.

\section*{\sf Clusters in X--rays}
The first $X$--ray observations of nearby galaxy clusters
\cite{bi:Kell,1xobs,bi:Giac74} showed that they are in general
associated with extended $X$--ray sources \cite{bi:cav71}, with
luminosities $L_X$ in the range $\sim 10^{43}$--$10^{45}$ erg
s$^{-1}$, whose emission originates in a diffuse hot intergalactic
medium permeating the cluster potential well (see \cite{bi:S88}, for a
historical review).  Since at equilibrium gas and galaxies in the
cluster have to share the same dynamics, one expects to measure a gas
temperature $k_BT\simeq \mu m_p \sigma_v^2$, where $m_p$ is the proton
mass and $\mu\simeq 0.6$ is the gas mean molecular weight.  Given the
typical cluster velocity dispersions, this corresponds to a
temperature of a few keV, which is what is indeed measured by \xray
observations.  At such energies the intra--cluster medium (ICM), which
is composed mainly by hydrogen, behaves like a fully ionised plasma
with an atomic density of $\sim 10^{-3}$ particles per cm$^{3}$.  The
scattering between free electrons and ions in such conditions produces
a {\sl thermal bremsstrahlung} radiation, which peaks in the \xray
region of the electromagnetic spectrum.  For this mechanism, the
emissivity (i.e., the energy released per unit time, frequency and
volume) at frequency $\nu$ and temperature $T$ scales as
$\epsilon_\nu\propto n_e n_i T^{-1/2}\exp{(-h_P\nu/k_BT)}$, where
$n_e$ and $n_i$ are the number densities of electrons and ions,
respectively, and $h_P$ is the Planck constant. This expression shows
why the identification of clusters in the $X$--ray band is less
affected by projection effects with respect to the optical selection
based on galaxy overdensities: while the optical emission depends
linearly on the number of galaxies, the $X$--ray emission depends on
the square of the local gas density, making clusters stand out more
sharply in the $X$--ray than in the optical light.  The contours in
Fig.~4 explicitly show how in the $X$--ray band the cluster emerges as
a single, practically isolated, extended source.

The $X$--ray luminosity of galaxy clusters is a
direct function of the total cluster mass, at least in a
phenomenological way \cite{Reiprich_Hans}, as demonstrated by the plot
in the bottom panel of Fig.~5. This is further supported by the
observation of a well--defined relation between the $X$--ray
luminosity and the temperature of the ICM, the latter being a reliable
indicator of the depth of the cluster gravitational potential
\cite{bi:emn96,bi:bn98}. Such a relationship has been observationally
calibrated at low redshift with a fairly large number of clusters,
showing that $L_{X}\propto T^\alpha$ with $\alpha\simeq 3$, a small
scatter, $\mincir 30\%$, around this relation \cite{bi:WJF97,bi:AE99}.
and no evidence for evolution out to $z\simeq 1.3$
\cite{bi:ms97,bi:rdc00,bi:H00,bi:rdcs2000}.  These observations have profound
implications for the physics of the ICM, as the steep slope of the
luminosity--temperature relation and the lack of evolution can not be
accounted for by the action of gravity only.  Additional physical
mechanisms, like radiative cooling in the central cluster regions and
heating from supernovae explosions and active galactic nuclei, seem to
play a key role in the thermodynamics of the intra--cluster gas (e.g.,
\cite{bi:TN00,bi:Bow00}, and references therein).  It is worth
emphasizing that, whichever mechanism determines the ICM
thermodynamics, the very fact that gas temperature and luminosity are
observed to be closely correlated demonstrates that $L_X$ is indeed a
reliable diagnostic of the cluster mass.

Besides $X$--ray imaging, radio observations of the
Sunyaev--Zel'dovich (SZ) effect are becoming an important alternative
method to detect distant clusters with a well--defined mass selection
(see \cite{bi:Bart} and \cite{bi:Carlstrom} for recent reviews). The
SZ effect is observed as 
a surface brightness variation in the Cosmic Microwave Background
(CMB) in the direction of a galaxy cluster and is produced by the
scattering of CMB photons on the energetic electrons of the ICM (what
is known as {\sl inverse Compton}). Owing to its nature, the SZ signal
depends only on the total thermal energy of the ICM, and is less
sensitive than the $X$--ray emission to its detailed structure and
complexity. Although the SZ signal has been now detected for several
known clusters, no systematic blind search based on this effect has
been realized to date.  The extensive use of this technique for
large--scale structure studies will become feasible only with the next
generation of high--resolution micro--wave surveyors, as the Planck
satellite.

\section*{\sf Clustering of X--ray clusters and implications for cosmology} 
A quantitative measure of the clustering of $X$--ray clusters has been
possible only in recent years, and on rather pioneeristic samples
\cite{lahav89}.  A major leap forward has been provided by the {\tt
ROSAT} All--Sky Survey (RASS, \cite{voges}), which for the first time
provided a $X$--ray imaging survey of the whole sky, in which
thousands of clusters as faint as $f_X=10^{-12}$erg~s$^{-1}$~cm$^{-2}$
can be detected
\footnote{Depending on the technology of their mirrors and detectors,
different $X$--ray satellites cover different energy ranges, which
need to be specified when quoting fluxes and luminosities.  The {\tt
ROSAT} band used here covers the energy range 0.1--2.4 keV. }.  Early
studies of $X$--ray clusters identified within sub--regions of the
RASS produced first estimates of the clustering of these objects
\cite{Romer94,nbr,bi:bpk99,bi:mosc00a}.  The quality of such $X$--ray
cluster catalogues has substantially improved through the recent
completion of the optical identification and redshift measurement for
a statistically complete sample of nearly 500 RASS clusters over half
of the sky \cite{Hans98,Gigi99} (see also Fig.~1). Benefitting of the
precision intrinsic to the $X$--ray selection, these data are now
providing an unprecedented large--scale description of the structure
of the Universe and a powerful testbed for cosmological
scenarios. Figs. 2 and 3 highlight the accuracy with which the
clustering of clusters is now measured on intermediate scales ($\sim
10$--$100\hm$) by the correlation function \cite{xi_REFLEX} and on the
largest scales achievable to date ($\sim 500\hm$) with the power
spectrum \cite{pk_REFLEX}.  Extension of this work to the whole sky
\cite{bi:noras} and to fainter fluxes will eventually result in a
complete sample of about 1500 clusters at a median redshift $z \simeq
0.1$.

Given the well--defined $X$--ray selection function for these samples,
the interpretation of the observed clustering in terms of cosmological
models is now relatively straightforward.  Once the gravitational
growth of cluster--sized mass clumps within a given model is
predicted, either analytically \cite{bi:MW96,bi:sheth} or numerically
\cite{bi:Gov99,bi:Virgo}, then the corresponding $X$--ray emission
from the ICM can be computed using the mass--temperature--luminosity
relation discussed in the previous section. Fig.~6 is an illustrative
example of the development of cosmic structures as generated by a
modern numerical N--body experiment, simulating the gravitational
growth of clustering from given initial conditions and capable to
precisely resolve individual cluster--sized clumps while at the same
time following their distribution on scales of several hundreds
Megaparsecs.  The mass--luminosity connection makes it possible to
compute the correlation function and power spectrum that clusters with a
given $X$--ray luminosity are predicted to have in each model
\cite{bi:bpk99,bi:mosc00a,pk_REFLEX,xi_REFLEX}.  Remarkably, the
amplitude of the correlation function and the scales over which
clusters are observed to be still inhomogeneously distributed,
consistently require a low--density Universe dominated by cold dark
matter (CDM), with density parameter $\Omega_m \sim 0.3$.  This is
illustrated by the two curves in Figs. 2 and 3, chosen to be both
consistent with CMB anisotropies on small \cite{bi:boom,bi:max} and
large \cite{bi:cobenorm} angular scales, but with different
$\Omega_m$.  Even a small increase of the density parameter from 0.3
to 0.5 produces a clear lack of coherence on scales above $100\hm$.
This gives an idea not only of the quality reached by current data,
but also of the role future surveys of $X$--ray clusters could play in
the framework of next decade high--precision cosmology, when
observations at different wavelengths will combine to pin down the
values of cosmological parameters with high accuracy
(e.g. \cite{CosmicTriangle}).

\section*{\sf Clusters and the evolution of the Universe}
We have marked in Fig.~6 the positions of clusters with a mass
corresponding to a temperature larger than 3 keV, which translates
into a luminosity of about $10^{44}\lum$. The two $z>0$ snapshots
correspond to look--back times of 5.7 and 9.0 billion years for the
L03 cosmology and to 6.6 and 9.5 billion years for the EdS cosmology.
Despite the similar pattern produced at the present time ($z=0$), the
past histories of the two models are very different. The most striking
feature is probably the fast decay in the abundance of hot, massive
clusters as a function of redshift in the $\Omega_m=1$ model, in
contrast to the mild changes visible in the low--density model. This
remarkable evolutionary difference represents one of the major
motivations for the recent deep \xray searches of clusters down to
fluxes about 1/100 that of the RASS (see refs.
\cite{bi:RosRev00,bi:GioiaRev00} and references therein).  Clusters at
$z\simeq 0.5$ are nowadays not considered as exceptions and even few
examples of $z\magcir 1$ $X$--ray bright clusters are now known
\cite{bi:Ros99}. The major result reached by these surveys is the
evidence for a weak evolution of the bulk of the cluster population
out to $z\simeq 1$, again consistent with the picture of a
low--$\Omega_m$ Universe.

Besides the cluster number density -- which basically measures the
degree of initial inhomogeneity on scales $\mincir 10\hm$ Mpc -- the
large--scale pattern also evolves in a markedly different way between
the two models of Fig.~6, this difference being mainly driven by the
value of the density parameter \cite{bi:Suto00,bi:mosc00b} (see Box
1).  Could we follow this evolution back in redshift using available
$X$--ray selected clusters?  Unfortunately, this is not yet possible:
even the largest deep surveys contain only $\sim 100$ clusters within
relatively small patches spread over the whole sky, which gives no
chance to map their distribution within a sufficiently large
contiguous volume at redshift $z>0.3$. For this to be possible, a
large--area $X$--ray survey would be needed, reaching a flux limit
significantly fainter than the {\tt ROSAT} All Sky Survey\footnote{
For example, detecting a typical cluster with $X$--ray luminosity
$L_X\sim10^{44}\; h^{-2}$ erg s$^{-1}$ at redshift $z\simeq 1$,
requires a survey limiting flux about 1/100 of that of the RASS.}.

A strong progress in $X$--ray astronomy is foreseen for the next few
years, in relation to the newly--launched AXAF-Chandra and XMM-Newton
$X$--ray satellites.  Also, new improved $X$--ray observatories are
currently under study \cite{satellites}. Still, none of these missions
will be ideal for extended studies of large--scale structure, as they
are not designed to perform an all--sky survey, or at least to cover a
large enough ($\sim 1000$ sq. deg.) contiguous area.  Covering the
whole sky in a reasonable time and with about 100 times higher
sensitivity than the {\tt ROSAT} All--Sky Survey would be feasible
even now, given current advances in the technology of $X$--ray optics
\cite{ref_specchi}. An all--sky survey to this depth would contain
more than 100,000 clusters out to $z\sim 1$, with about 15,000 of
these laying above $z\magcir 0.5$. Although no such experiment is
currently funded, several ideas and proposals are circulating in the
scientific community. We can hope that the enormous potential of
$X$--ray clusters to follow the shaping process of the Universe will
soon make these projects become reality.

\section*{\sf Acknowledgments}
We thank the REFLEX collaboration for giving us the possibility to
discuss material in advance of publication; Chris Collins, Fabio
Governato and Meg Urry for reading the manuscript; Davide Lazzati and
Alberto Moretti for their help with some of the figures; and Marisa
Girardi for providing the data files on which Fig.~5 is based. We also
acknowledge several enlightening discussions on \xray clusters with
Piero Rosati.

\section*{Box 1: A large--scale structure primer}
The {\sl redshift} $z$ of a galaxy is defined as the fractional
increase in the observed wavelength of the emitted radiation with
respect to its laboratory value: $z=(\lambda - \lambda_o)/\lambda_o$.
In the standard cosmological model this is interpreted as due to the
global expansion of the Universe, and to first approximation (i.e.,
neglecting any acceleration of the cosmic expansion) it can be thought
as the consequence of a recession velocity $v_{rec} = c\,z$ (where $c$
is the speed of light), proportional to the distance to the galaxy
itself, $v_{rec} \simeq H_o\,d$. This is the empirically verified {\sl
Hubble law} and the constant of proportionality $H_o$ is the famous
{\sl Hubble constant}.  Through this relation, we can measure the
distance to far--away galaxies by simply observing the red--shift in
their spectrum.

According to the picture of gravitational instability, cosmic
structures arise from the gravitational growth on initially tiny
inhomogeneities, generated in the very early stages of the Universe
life. These cosmic inhomogeneities are described by the fluctuation
field $\delta(\vec x,z)\,=\,\left(\delta\rho / \rho\right)_{\vec
x,z}$, where $ \rho(\vec x,z)$ is the density of the Universe at the
point $\vec x$ and at redshift $z$.  A basic statistical
characterization of $\delta(\vec x)$ is represented by its two--point
correlation function, $\xi(r)$, which describes the excess
fluctuations with respect to a uniform distribution.  The Fourier
transform of $\xi(r)$ is the power spectrum
\begin{equation}
P(k)\,=\,{\frac{1}{2\pi^2}}\int_0^\infty dr\,\xi(r)\,{\frac{\sin kr}{kr}}\,.
\end{equation}
The power spectra of two representative cosmological models are
compared in Fig.~2 to observational results.  Even keeping fixed the
assumption of a Universe dominated by Cold Dark Matter (CDM),
different shapes for the matter $P(k)$ can be obtained by varying the
values of $\Omega_m$, $h$ and the amount of baryons \cite{bi:eh98}.
Its amplitude is instead usually determined in a phenomenological way,
by matching some observed measure of the root--mean--squared ({\sl
rms}) fluctuation (in the mass, not in the light!) at some scale.  One
way is to reproduce the level of CMB anisotropy measured by the {\sl
COBE} satellite \cite{bi:cobenorm}, that probes fluctuations on scales
$\sim 10^3$Mpc.  On a completely different range of scales, an
alternative method is offered by the local abundance of galaxy
clusters, in virtue of their ability to probe the mass fluctuations.
Since they formed from the collapse of density fluctuations on a scale
$\simeq 10\,h^{-1}{\rm Mpc}$, their number density today is directly
proportional to the {\sl rms} amplitude of such fluctuations at the
onset of their growth.  As massive clusters arise from rare high peaks
of the density fluctuation field, their number density is rather
sensitive to the amplitude of these fluctuations, providing a precise
constraint on the amplitude of $P(k)$
\cite{bi:Eke96,bi:G98b,bi:VL99,CosmicTriangle}. For instance, the
simulations shown in Fig. 6 refer to two models having different
$\Omega_m$ but normalisation tuned so as to produce a comparable
number of clusters at $z=0$.

Once $P(k)$ is normalised at the present time, its amplitude in the
past depends on the evolution of density fluctuations.  As long as
fluctuations are small, their growth is independent of the position,
so that $\delta(\vec x, z)=D(z)\delta(\vec x,z=0)$.  The redshift
dependence of $D(z)$ is determined by comparing the time--scales for
gravitational collapse of a perturbation and for the expansion of the
Universe \cite{bi:JAP_book,bi:CL95_book}. These two time scales are
actually always identical when $\Omega_m=1$, thus implying that the
perturbation growth equals the cosmic expansion factor:
$D(z)=(1+z)^{-1}$. For a low--density Universe, the two time--scales
are similar at early epochs, while the cosmic expansion overtakes the
fluctuation growth when the former stops feeling the self--gravity of
the Universe. This happens at an epoch corresponding to $1+z\simeq
\Omega_m^{-1/3}$ or $1+z\simeq \Omega_m^{-1}$ in the cases with and
without a cosmological constant providing flat geometry,
respectively. Therefore, in a low--$\Omega_m$ Universe the
fluctuations growth progressively slows down at later epochs and
eventually freezes.

The plots in Fig.~6 clearly show the different degree of evolution
characterising models with different values of $\Omega_m$, as
witnessed by the different number of clusters already formed at high
redshift. For this reason, the determination of the number density of
distant clusters is a powerful diagnostic for cosmological models (see
refs.\cite{bi:BRTN99,bi:H00} and references therein).

\section*{Figures}
\subsection*{Figure 1:}
Maps of the cosmic web. The top panel shows the distribution of $\sim
26,000$ galaxies in the Northern and Southern slices of the Las
Campanas Redshift Survey\cite{lcrs} (LCRS), to a maximum distance of
$600\hm$ from the observer, located in the centre.  The typical
morphology of large--scale structure, with filaments and voids is
evident.  Superimposed on the ``blue dust'' of the galaxy
distribution, the green circles mark the positions of those $X$--ray
clusters of galaxies from the REFLEX survey \cite{Hans98} that lie
approximately within the volume of the LCRS.  The full volume of the
REFLEX survey (which contains all clusters brighter than an $X$--ray
flux of $3\times 10^{-12}\fl$ over a large part of the southern sky)
within the same distance limit, is shown in the bottom panel (similar
orientation, South is up for clarity).  The much larger volume sampled
by clusters and their clustering along filamentary structures is
evident.  The missing part of the hemisphere corresponds to the region
highly obscured by the Milky Way disk ($\pm 20^\circ$ in galactic
latitude).  Similar volumes will be partly filled by galaxy redshift
measurements only in the coming years \cite{Margon,Colless}.

\subsection*{Figure 2:}
Statistical description of clustering. The two--point correlation
functions $\xi$ of galaxies (squares) and $X$--ray clusters of
galaxies (circles), computed from the two data sets of Fig.~1
\cite{xi_lcrs,xi_REFLEX}, plotted as a function of separation $r_s$
(where the suffix $s$ indicates that all object distances are computed
from the measured redshift). The curves are the predictions of two CDM
models, with different density parameters $\Omega_m$, for an $X$--ray
cluster survey with the same flux--limit as the real data.  Solid
curve: $\Omega_m=0.3$ and Hubble parameter $h=0.7$; dashed curve:
$\Omega_m=0.5$ and $h=0.6$. Both models have flat spatial geometry
provided by a cosmological constant contribution
(i.e. $\Omega_m+\Omega_\Lambda=1$) and power--spectrum normalization
chosen so as to be consistent measured CMB anisotropies
\cite{bi:cobenorm,bi:boom,bi:max}. Also, we take
$\Omega_{bar}h^2=0.019$ for the baryon density \cite{bi:tytl}.

\subsection*{Figure 3:}
The power spectrum of the distribution of galaxies and $X$--ray
clusters of galaxies from the data of Fig.~1.  The squares are from
galaxy data, which in addition to the LCRS points (filled squares
\cite{LCRS_pk}), include a measure from another survey with better
volume coverage (open squares \cite{Durham-UKST}). The filled circles
show instead an estimate of the power spectrum of $X$-ray selected
clusters from the REFLEX survey \cite{pk_REFLEX}.  Note the rather
different amplitude between the galaxy and cluster power spectra,
similarly to that shown by correlation functions (see Fig.~2). The two
curves are theoretical predictions from the same cosmological models
shown in Fig.~2.

\subsection*{Figure 4:}
The visual and $X$--ray appearance of the rich galaxy cluster
RXCJ1206.2-0848 at redshift $z=0.44$, discovered by the REFLEX survey
\cite{Gigi99}.  The optical image has been obtained by combining three
images taken with the ESO 3.6~m telescope, corresponding to the red,
green and blue bands.  The yellowish colour of objects in the central
region reflects the old stellar population characterising elliptical
galaxies, that typically dominate in cluster cores.  Note also the
presence of at least two probable gravitational arcs (with blue
colour), near the central giant elliptical galaxy.  These are the
result of the {\sl gravitational lensing} phenomenon, by which the
cluster mass distorts and amplifies the images of background galaxies
(thus providing an independent way to probe its potential
\cite{Mellier}).  The contours show the $X$--ray emission from the
cluster measured by the {\tt ROSAT} All--Sky Survey, which yield a
luminosity $L_X\simeq 5\times 10^{44}\; h^{-2}$ erg s$^{-1}$.  The
field of view in this image is about 5 arcmin, corresponding to a
physical size of $\sim 1\hm$ at the cluster distance.

\subsection*{Figure 5:}
Correlations between mass and observational properties of galaxy
clusters.  The correlations with Abell \cite{bi:A58,bi:ACO89} richness
counts $N_c$ (upper panel) and their $X$--ray bolometric luminosity
$L_{X,bol}$ (lower panel), based on the merge of a compilation of
clusters with accurate measures of velocity dispersion \cite{bi:G98}
and a sample of $X$--ray bright Abell clusters (XBACs
\cite{bi:xbacs}). Cluster masses are estimated by applying the virial
theorem to the velocity dispersions of member galaxies. In both
panels, filled circles are for those clusters belonging to XBACs,
while open circles are for those Abell clusters that have too low a
$X$--ray emission to be included in XBACs.  The weak correlation in
the upper panel demonstrates the difficulty of using clusters selected
by optical richness to strongly constrain cosmological scenarios. This
aim is better achieved with clusters selected by $X$--ray luminosity,
which clearly shows a tighter correlation with the cluster mass.

\subsection*{Figure 6:}
The evolution of gravitational clustering simulated using an N--body
code for two different models.  Each of the three redshift snapshots
shows a region with 250$\hm$ side and 75$\,h^{-1}{\rm Mpc}$ thick
(co--moving with the cosmic expansion).  Each simulation contains
about two millions particles.  The upper one describes a flat
low--density model with $\Omega_m=0.3$ and $\Omega_\Lambda=0.7$ (L03),
while the lower an Einstein--de-Sitter model (EdS) with $\Omega_m=1$.
In both cases the amplitude of the power spectrum is consistent with
the number density of nearby galaxy clusters \cite{bi:Eke96,bi:G98b}
and with the large--scale CMB anisotropies \cite{bi:cobenorm}.
Superimposed on the dark matter distribution, the yellow circles mark
the positions of galaxy clusters that would be seen shining in
$X$--rays with a temperature $T>3$ keV, as computed from the cluster
mass according to the relation calibrated from hydrodynamical
simulations \cite{bi:emn96}. The size of the circles is proportional
to temperature.

\end{document}